\documentclass{emulateapj}
\usepackage{apjfonts}

\usepackage{amsmath}

\usepackage{graphics}

\newcommand{\bd}{\begin{displaymath}}
\newcommand{\ed}{\end{displaymath}}

\slugcomment{Received 2008 December 17; accepted 2009 February 12}

\shorttitle{On the BL Lac objects/radio quasars and the FRI/II
dichotomy} \shortauthors{Xu Y.D. et al.}

\begin{document}

\title{On the BL Lacertae objects/radio quasars and the FRI/II dichotomy}

\author{Ya-Di Xu$^{1}$, Xinwu Cao$^{2}$, Qingwen Wu$^{3}$}
\affil{$^{1}$Physics Department, Shanghai Jiao Tong University,800
Dongchuan Road, Min Hang, Shanghai, 200240, China}

\affil{$^{2}$Key Laboratory for Research in Galaxies and Cosmology,
Shanghai Astronomical Observatory, Chinese Academy of Sciences, 80
Nandan Road, Shanghai 200030, China}

\affil{$^{3}$International Center for Astrophysics, Korean Astronomy
and Space Science Institute, Daejeon 305348, Republic of Korean\\
Email: ydxu@sjtu.edu.cn, cxw@shao.ac.cn,qwwu@shao.ac.cn. }

\clearpage

\begin{abstract}

In the frame of unification schemes for radio-loud active galactic
nuclei (AGNs), FR I radio galaxies are believed to be BL Lacertae
(BL Lac) objects  with the relativistic jet misaligned to our line
of sight, and FR II radio galaxies correspond to misaligned radio
quasars. The Ledlow-Owen dividing line for FR I/FR II dichotomy in
the optical absolute magnitude of host galaxy--radio luminosity
($M_R$--$L_{\rm Rad}$) plane can be translated to the line in the
black hole mass--jet power ($M_{\rm bh}$--$Q_{\rm jet}$) plane by
using two empirical relations: $Q_{\rm jet}$--$L_{\rm Rad}$ and
$M_{\rm bh}$--$M_R$. We use a sample of radio quasars and BL Lac
objects with measured black hole masses to explore the relation of
the jet power with black hole mass, in which the jet power is
estimated from the extended radio emission. It is found that the BL
Lac objects are clearly separated from radio quasars by the Ledlow
\& Owen FR I/II dividing line in the $M_{\rm bh}$--$Q_{\rm jet}$
plane. This strongly supports the unification schemes for FR I/BL
Lac object and FR II/radio quasar. We find that the Eddington ratios
$L_{\rm bol}/L_{\rm Edd}$ of BL Lac objects are systematically lower
than those of radio quasars in the sample with a rough division at
$L_{\rm bol}/L_{\rm Edd}\sim 0.01$, and the distribution of
Eddington ratios of BL Lac objects/quasars exhibits a bimodal
nature, which imply that the accretion mode of BL Lac objects may be
different from that of radio quasars.
\end{abstract}

\keywords{black hole physics
---galaxies: active---galaxies: nuclei---quasars: emission lines---BL Lacertae objects: general}

\section{Introduction}

FR I radio galaxies (defined by edge-darkened radio structure) have
lower radio power than FR II galaxies (defined by edge-brightened
radio structure due to compact jet-terminating hot spots)
\citep{1974MNRAS.167P..31F}. Relativistic jets are observed in many
radio-loud active galactic nuclei (AGNs). In the frame of
unification schemes of radio-loud AGNs, FR I radio galaxies are
believed to be misaligned BL Lacertae (BL Lac) objects, and FR II
radio galaxies correspond to misaligned radio quasars
\citep*[see][for a review]{1995PASP..107..803U}. Most BL Lac objects
have featureless optical and ultraviolet continuum spectra, and only
a small fraction of BL Lac objects show very weak broad emission
lines, while quasars usually have strong broad-line emission.  The
broad emission lines of quasars are produced by distant gas clouds
in broad-line regions (BLR), which are photo-ionized by the
optical/UV continua radiated from the accretion disks surrounding
massive black holes. The difference of the broad-line emission
between radio-loud quasars and BL Lac objects may be attributed to
their different central engines
\citep*[e.g.,][]{2002ApJ...571..226C,2002ApJ...570L..13C,2003ApJ...599..147C}.

The unified scheme of BL Lac objects and FR I radio galaxies have
been extensively explored by many previous authors with different
approaches, such as the comparisons of the spectral energy
distributions (SEDs) in different wavebands
\citep*[e.g.,][]{1996AJ....111...53O,2000MNRAS.318..493C,2001ApJ...548..244B},
the radio morphology, {the radio luminosity functions (LFs)
\citep*[e.g.,][]{1991ApJ...368..373P,1992AJ....104.1687K,1993AJ....106..875L}
and the optical line emission}
\citep*[e.g.,][]{2005MNRAS.361..469M}. \citet{1992ApJ...387..449P}
derived the radio LFs of flat spectrum radio quasars (FSRQs) and FR
II galaxies from a sample of radio-loud AGNs. They considered a
two-component model, in which the total luminosity is the sum of an
unbeamed part and a beamed jet luminosity. {The beamed LFs of FR II
radio galaxies are consistent with the observed LFs of FSRQs and
steep spectrum radio quasars (SSRQs)}, which strengthens the
unification of FR II galaxies and radio quasars \citep*[see][for the
details]{1992ApJ...387..449P}. Similar analyses were carried out on
the relation between FR I galaxies and BL Lac objects
\citep{1991ApJ...368..373P,1995PASP..107..803U}, which is also
consistent with the unification of FR Is and BL Lac objects. { Even
though the main observational features of different types of
radio-loud AGNs can be successfully explained in the frame of the
unification schemes, some authors have found observations indicating
that the unification may be more complex than usually portrayed in
these schemes}
\citep*[e.g.,][]{2005MNRAS.361..469M,2008MNRAS.391..967L}.
\citet{2008MNRAS.391..967L} found that a considerable number of BL
Lac objects can be identified with the relativistically beamed
counterparts of FR II radio galaxies in a sample of BL Lac objects
selected from the Deep X-ray Radio Blazar Survey (DXRBS).

\citet{1996AJ....112....9L} found that FR I and FR II radio galaxies
can be clearly divided in the host galaxy optical luminosity--radio
luminosity ($M_R$--$L_{\rm Rad}$) plane, by a dividing line showing
that radio power is proportional to the optical luminosity of the
host galaxy. What causes the FR I/FR II division is still unclear,
{and there are two categories of models to explain it: (1) the
morphological differences being caused by the interaction of the
jets with the ambient medium of different physical properties
\citep* [e.g.,][]{2000A&A...363..507G}; and/or (2) different
intrinsic nuclear properties of accretion and jet formation
processes \citep* [e.g.,][]{1995ApJ...451...88B,
1995ApJS..101...29B, 1996MNRAS.283L.111R, 2001A&A...379L...1G,
2004MNRAS.351..733M, 2007MNRAS.376.1849H}}.
\citet{2001A&A...379L...1G} used the optical luminosity of the host
galaxy to estimate the central black hole masses of FR I/FR II radio
galaxies, and the bolometric luminosity is estimated from the
{radio} power of jets in FR I/FR II galaxies. They suggested that
most FR I radio galaxies are accreting at lower rates compared with
FR IIs, {which could correspond to different accretion modes in FR I
and FR II radio galaxies.} {If the black hole is spinning rapidly,
the rotational energy of the black hole is expected to be
transferred to the jets by the magnetic fields threading the holes,
namely,  the Blandford-Znajek (BZ) mechanism
\citep{1977MNRAS.179..433B}. The jet can also be accelerated by the
large-scale fields threading the rotating accretion disk
\citep*[i.e., the Blandford-Payne (BP)
mechanism,][]{1982MNRAS.199..883B}. \citet{2004MNRAS.349.1419C}
found that the BZ mechanism for rapidly spinning black holes
surrounded by advection dominated accretion flows (ADAFs)
\citep{1995ApJ...452..710N} provides insufficient power to explain
the jets in some 3CR FR I radio galaxies.}
\citet{2008ApJ...687..156W} calculated the maximal jet power
{available from ADAFs around Kerr black holes} as a function of
black hole mass with an hybrid jet formation model (i.e., BP+BZ
mechanism). {They found that it can roughly reproduce the dividing
line of the Ledlow-Owen relation for FR I/FR II dichotomy in the
black hole mass--jet power ($M_{\rm bh}$--$Q_{\rm jet}$) plane with
the mass accretion rate $\dot{M}\sim 0.01\dot{M}_{\rm Edd}$, if the
black hole spin parameter $a\sim 0.9-0.99$ is adopted. This
accretion rate indicates that FR I and FR II galaxies have different
accretion modes, supporting the results of
\citet{2001A&A...379L...1G} and suggesting that FR I sources are in
the ADAF mode.} \citet{2008ApJ...687..156W}'s results imply that the
black hole spin may play an important role in the jet formation at
least for FR I radio galaxies \citep*[also see][for the discussion
of the impact of black hole spin on the jet formation in
AGNs]{2007ApJ...658..815S}.

In this work, we use a sample of BL Lac objects and radio quasars
with measured radio power, black hole masses, and Eddington ratios,
to explore the relationship between BL Lac objects and radio
quasars, and to compare it with the FR I/FR II division.  The sample
and the estimates of black hole mass/jet power are described in \S 2
and \S 3. We show the results in \S 4, and \S 5 contains the
discussion. The cosmological parameters $H_{0}=70~\rm km~s^{-1}
Mpc^{-1}$, $\Omega_{M} = 0.3$ and $\Omega_{\Lambda} = 0.7$ have been
adopted in this work.

\section{The sample}

The host galaxies of 132 BL Lac objects {have been observed} with
the {\it Hubble Space Telescope} WFPC2 by
\citet{2000ApJ...532..816U}, among which there are 48 sources with
measured redshifts and extended radio emission. We add additional 18
BL Lac objects compiled in the work of \citet{2008astro-ph/08041180}
to the \citet{2000ApJ...532..816U}'s sample, which leads to 66 BL
Lac objects ({including 28 low-energy-peaked BL Lac objects (LBLs)
and 38 high-energy-peaked BL Lac objects (HBLs)}) with measured
redshifts and extended radio emission data for our present
investigation. We search the literature for the  emission line data
of these sources, {and find 44 sources including 23 LBLs and 21
HBLs.} We use the luminosity of narrow line {[O\,{\sc ii}]} at
3727~$\rm \AA$ to estimate the bolometric luminosity. {For the
sources in which the emission line data of {[O\,{\sc ii}]} being
unavailable, we estimate the {[O\,{\sc ii}]} luminosity using other
narrow emission lines.}

In order to compare the difference between BL Lac objects and radio
quasars, we need a sample of radio quasars. In this work, we adopt
the sample of radio quasars compiled by \citet{2006ApJ...637..669L},
which is selected from the 1 Jy, S4, and S5 radio source catalogs.
Their sample consists of 146 radio quasars including 79 FSRQs (with
$\alpha_{2-8\rm GHz}<0.5$) and 67 steep-spectrum radio quasars
(SSRQs) (with $\alpha_{2-8\rm GHz}>0.5$). All quasars in their
sample have estimated black hole masses and jet power
\citep*[see][for the details of the quasar
sample]{2006ApJ...637..669L}.

\section{The black hole mass and jet power}

{The relation between black hole mass $M_{\rm bh}$ and host galaxy
luminosity $L_{K}$ at $K$-band \citep*[Eq. 1
in][]{2004MNRAS.352.1390M} is derived from $M_{\rm bh}$--$M_{R}$ by
using an average color correction of $R-K=2.7$ for the same
cosmology adopted in this paper. We convert this relation back to
$M_{\rm bh}$--$M_{R}$ as }
\begin{equation}
\log_{10}(M_{\rm bh}/M_{\odot})=-0.50(\pm 0.02)M_{R}-2.75(\pm 0.53),
\label{eqmbh}
\end{equation}
to estimate the central black hole masses of BL Lac objects in this
sample. {For a few BL Lac objects, their black hole masses can also
be estimated from their stellar dispersion velocity $\sigma$ with
the empirical $M_{\rm bh}$--$\sigma$ relation.
It is found that the black hole masses of three BL Lac objects
estimated with $M_{\rm bh}$--$\sigma$ relation are roughly
consistent with those estimated with Eq. (\ref{eqmbh})
\citep*[see][for the details, and references
therein]{2004ApJ...609...80C}.}

The jet power can be estimated with the relation  between jet power
and radio luminosity proposed by \citet{1999MNRAS.309.1017W},
\begin{equation}
Q_{\rm jet}\simeq 3\times 10^{38}f^{3/2}L^{6/7} _{\rm ext,151}~~(\rm
W), \label{eqqjet}
\end{equation}
where $L_{\rm ext,151}$ is the extended radio luminosity at 151 MHz
in units of 10$^{28}$ W~Hz$^{-1}$~sr$^{-1}$.
\citet{1999MNRAS.309.1017W} have argued that the normalization is
uncertain and introduced the factor $f$ ($1\leq f\leq 20$) to
account for these uncertainties. This relation was proposed for FR
II radio galaxies and quasars. {\citet{2004MNRAS.349.1419C} compared
the power of the jet in M87 (a typical FR I radio galaxy) derived
with different approaches, and found that Eq. (\ref{eqqjet}) may
probably be suitable even for FR Is \citep*[see][for the details,
and references therein]{2004MNRAS.349.1419C}.} {Following
\citet{2003ApJ...599..147C},} we adopt this relation to estimate the
power of jets in BL Lac objects, which is believed to be a good
approximation {if BL Lacs can be unified with FR Is.}

For most BL Lac objects, their radio/optical continuum emission is
strongly beamed to us due to their relativistic jets and small
viewing angles of the jets with respect to the line of sight
\citep*[e.g.,][]{2003A&A...407..899F,2006A&A...450...39G}. The
low-frequency radio emission (e.g. 151 MHz) may still be Doppler
beamed. We therefore use the extended radio emission detected by VLA
to estimate the jet power, as adopted in
\citet{2003ApJ...599..147C}. The observed extended radio emission is
$K$-corrected to 151 MHz in the rest frame of the source assuming
$\alpha_{\rm e}=0.8$ ($f_{\nu}\propto \nu^{-\alpha_{\rm e}}$)
\citep{1999A&AS..139..601C}.

{We take the black hole masses of radio quasars from
\citet{2006ApJ...637..669L}, which are estimated from the broad line
widths of ${\rm H\,{\beta}}$, {Mg\,{\sc ii}}, or {C\,{\sc iv}}, as
well as the line luminosities of these lines \citep*[see][for the
details]{2006ApJ...637..669L}.} In \citet{2006ApJ...637..669L}'s
work, the jet power is estimated from the extended radio emission at
151~MHz with the formula derived by \citet{2005ApJ...623L...9P},
which is slightly different from Eq. (\ref{eqqjet}) proposed by
\citet{1999MNRAS.309.1017W}. To be self-consistent, we estimate the
jet power of quasars in \citet{2006ApJ...637..669L}'s sample from
their extended radio luminosities with Eq. (\ref{eqqjet}), which is
the same as the estimates of jet power for BL Lac objects in this
work.

For BL Lac objects, the observed optical continuum emission may be
dominated by the beamed synchrotron emission from the relativistic
jets \citep*[e.g.,][]{2006A&A...450...39G}. The narrow-line regions
are believed to be photo-ionized by the radiation from the accretion
disk, and the narrow-line emission can be used to estimate the
bolometric luminosity for BL Lac objects. We convert the luminosity
of the narrow-line {[O\,{\sc ii}]} to bolometric luminosity using
the relation proposed by \citet{1999MNRAS.309.1017W},
\begin{equation}
L_{\rm bol}=5\times 10^{3}L_{\rm [O\,{II}]}~{\rm W}, \label{eqlbol}
\end{equation}
for the BL Lac objects in this sample. For the objects which lack
{[O\,{\sc ii}]} line emission data, we convert the luminosities of
other narrow lines ({[O\,{\sc iii}]} or H$_\alpha$+{[N\,{\sc ii}]})
to the luminosity of {[O\,{\sc ii}]} using the ratios suggested by
\citet{1995ApJ...448..521Z} for FR I galaxies.  The narrow-line
emission data for the BL Lac objects are taken from the literature
\citep{2006A&A...457...35S,2003A&A...412..651C,
2000AJ....120.1626R,2001AJ....122..565R,1993A&AS...98..393S,
1996MNRAS.281..425M,1992MNRAS.254..546M}. We note that Eq.
(\ref{eqlbol}) is derived for FR IIs/quasars, while ADAFs may be
present in these BL Lac objects. The SED of an ADAF is significantly
different from that of a standard thin disk
\citep*[e.g.,][]{1995Natur.374..623N}. \citet{2002ApJ...567...73N}
calculated the emission of narrow-line regions photo-ionized by two
different SED templates respectively, i.e., a standard thin disk SED
template with a bump in UV/soft X-ray bands and a hot ADAF SED
template described by a power-law continuum in hard X-ray bands with
an exponential cutoff. They found that the narrow-line regions are
more efficiently photo-ionized by the ADAF SED template than the
standard thin disk case \citep*[see the bottom panel of Fig. 5
in][]{2002ApJ...567...73N}, which implies that the present estimates
on the bolometric luminosity with Eq. (\ref{eqlbol}) may be
over-estimated to some extent (a factor of $\sim 2-3$ for the
narrow-line regions with hydrogen column density $\la 10^{20}~{\rm
cm^{-2}}$).

For radio quasars, we estimate their bolometric luminosities from
the total broad-line luminosities {$L_{\rm BLR}$ calculated by
\citet{2006ApJ...637..669L}, as the optical continua for most
radio-loud quasars may probably be contaminated by the beamed
emission from relativistic jets. \citet{2006ApJ...637..669L} derived
a tight correlation: $\lambda L_{\lambda}(5100{\rm \AA)}=84.3L_{{\rm
H}\beta}^{0.998}$, for the sample of radio-quiet AGNs in
\citet{2000ApJ...533..631K}. Given that the luminosity of the
broad-line ${\rm H}_\beta$ corresponds to $\sim$4 per cent of
$L_{\rm BLR}$ \citep*[see][and references
therein]{2006ApJ...637..669L} and using the relation $L_{\rm
bol}\simeq 9\lambda L_{\lambda}(5100{\rm\AA)}$
\citep{2000ApJ...533..631K}, the bolometric luminosity can be
estimated as $L_{\rm bol}\simeq 30L_{\rm BLR}$.}

\section{The results}

The division between FR I and FR II radio galaxies is clearly shown
by a line in the plane of total radio luminosity and optical
luminosity of the host galaxy \citep{1996AJ....112....9L}. The
optical luminosity of the host galaxy can be converted to black hole
mass $M_{\rm bh}$ by using the empirical relation (\ref{eqmbh}),
while the jet power $Q_{\rm jet}$ can be estimated from the radio
luminosity with relation (\ref{eqqjet}). Thus, the dividing line
between FR I and II radio galaxies is translated to
\begin{equation}
\log Q_{\rm jet}({\rm ergs~ s^{-1}})=1.13 \log M_{\rm bh}(M_{\odot})
+33.18 +1.50\log f, \label{fri_ii}
\end{equation}
in $M_{\rm bh}$--$Q_{\rm jet}$ plane \citep*[see][for the
details]{2008ApJ...687..156W}, {which is modified for the cosmology
adopted in this paper.} In Figure \ref{qjetmbh}, we plot the
relation between the black hole masses $M_{\rm bh}$ and jet power
$Q_{\rm jet}$ for radio quasars and BL Lac objects. It is found that
BL Lac objects can be roughly separated from quasars by the FR I/II
dividing line.

The distributions of Eddington ratios  for BL Lac objects and
quasars are plotted in Fig. \ref{histlbol_all}, where only the BL
Lac objects with measured line emission have been included, because
the bolometric luminosity is derived from the  emission lines for
these sources. We estimate the statistical significance of a
possible bimodal distribution of Eddington ratios  for BL Lac
objects and quasars using the KMM algorithm
\citep{1994AJ....108.2348A}. The distribution for the entire sample
is strongly inconsistent with being unimodal ($P$-value$<0.001$),
and the KMM algorithm separates the entire sample into two groups.
No significant difference is found in the distributions of black
hole masses  for BL Lac objects and radio quasars.

\vskip 1.0cm \figurenum{1}
\centerline{\includegraphics[angle=0,width=9.0cm]{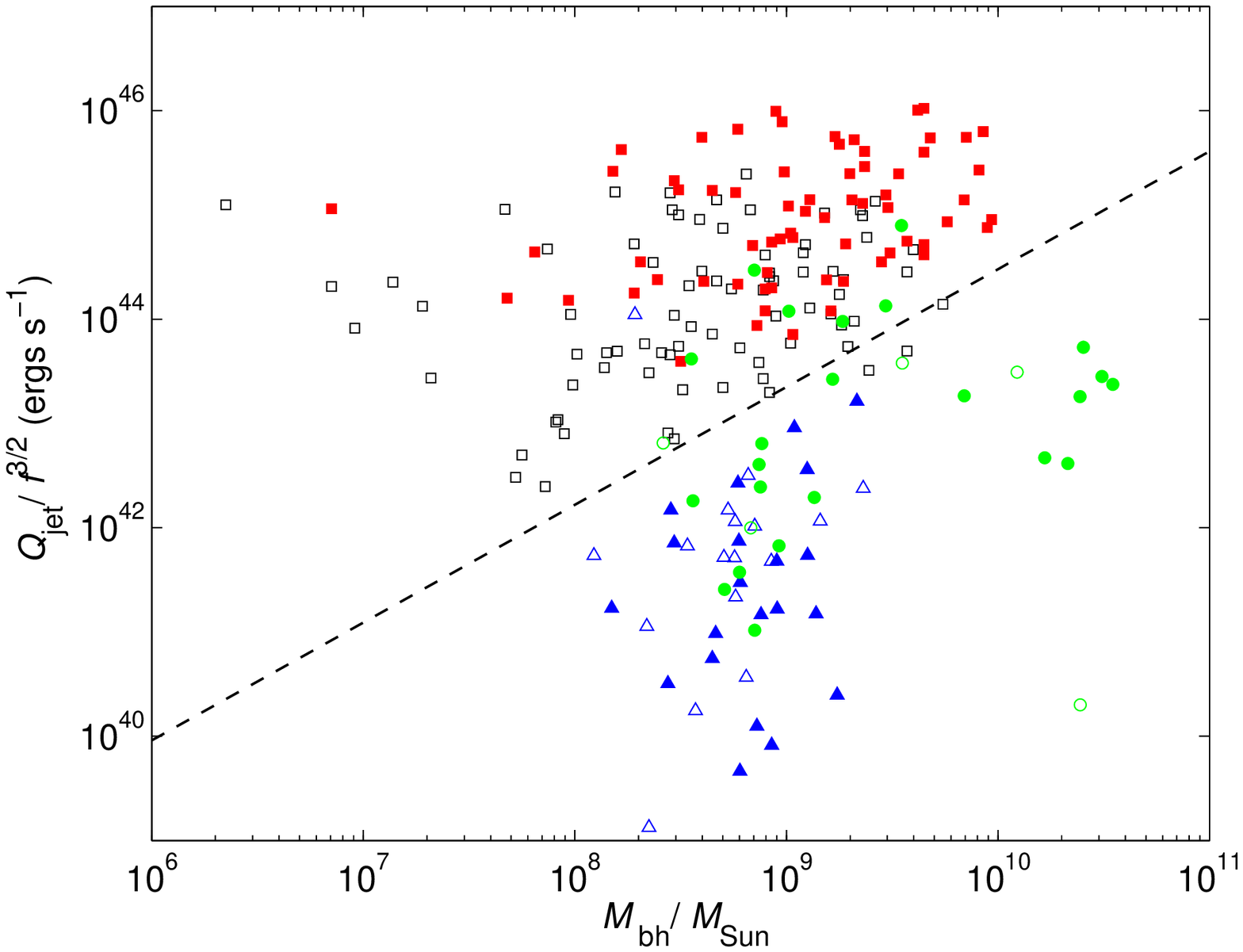}}
\figcaption{The relation between black hole mass $M_{\rm bh}$ and
jet power $Q_{\rm jet}$ for the BL Lac objects and quasars. The open
squares and filled squares represent FSRQs and SSRQs respectively,
while the circles and triangles represent BL Lac objects. The filled
circles/triangles represent the LBLs/HBLs with measured line
emission, while the open circles/triangles the LBLs/HBLs without
measured line emission. The dashed line represents the Ledlow-Owen
dividing line between FR I and FR II radio galaxies given by Eq.
(\ref{fri_ii}).} \label{qjetmbh} \centerline{}

\vskip 1.0cm \figurenum{2}
\centerline{\includegraphics[angle=0,width=9.0cm]{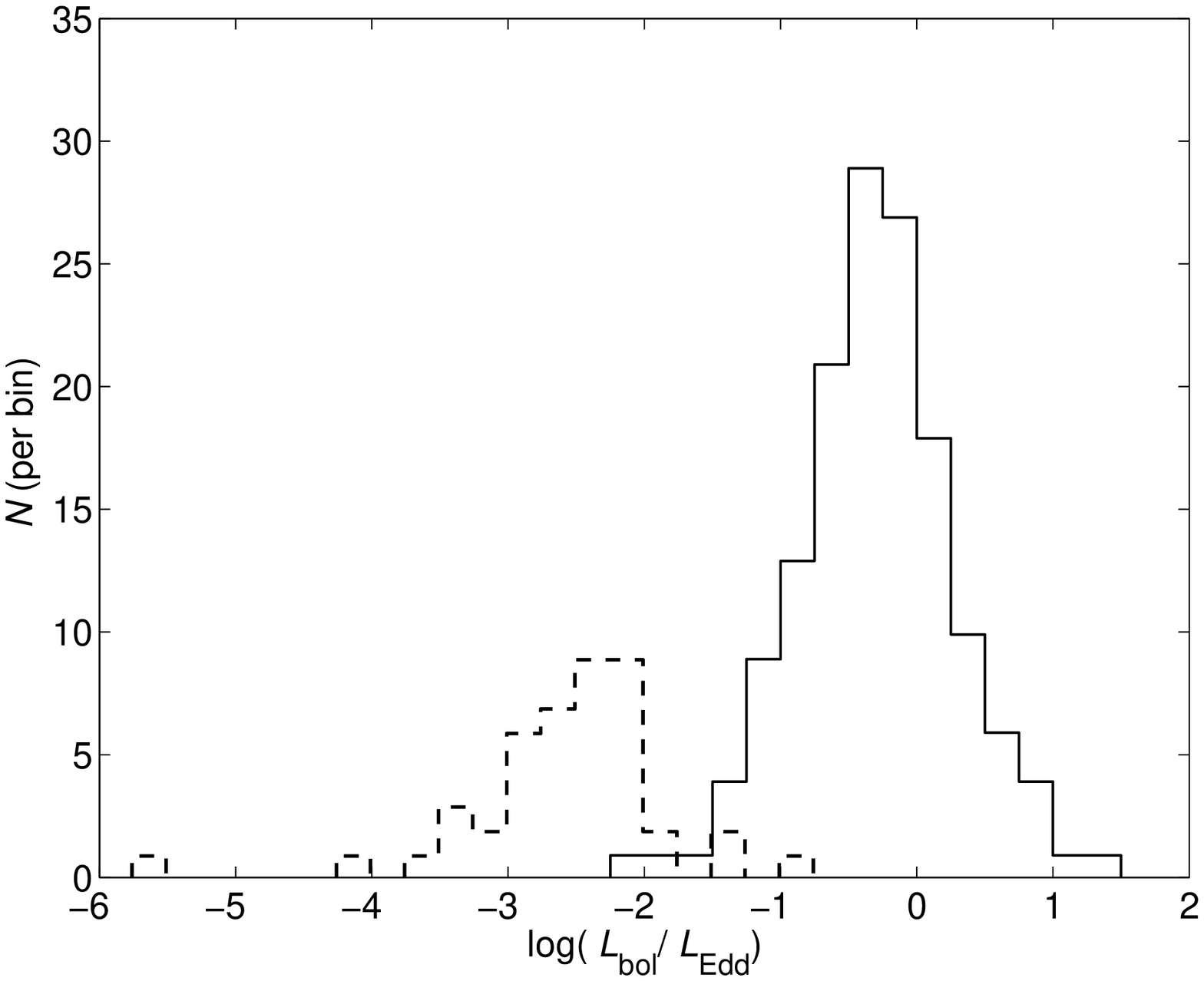}}
\figcaption{The distributions of Eddington ratios ($L_{\rm
bol}/L_{\rm Edd}$)  for BL Lac objects (dashed line) and quasars
(solid line), respectively. } \label{histlbol_all}\centerline{}

\section{Discussion}

Figure \ref{qjetmbh} shows that the FR I/FR II dividing line given
by \citet{1996AJ....112....9L} roughly separates the radio-loud
quasars from BL Lac objects in the $M_{\rm bh}$--$Q_{\rm jet}$
plane, which strongly supports the FR I/BL Lac objects and FR
II/radio quasars unification schemes. This conclusion is independent
of the value of the uncertainty factor $f$ in Eq. (\ref{eqqjet}).

We find that only a small fraction of LBLs/quasars are above/below
the dividing line, which is similar to the FR I/II division
\citep*[see for instance the Fig. 1 in][]{1996AJ....112....9L}. The
HBLs have relatively lower jet power than LBLs, and only one HBL
appears above the dividing line. This means that the BL Lacs/quasars
and the FR I/II divisions  may be true only in a statistical sense.
The exceptional sources in the $M_{\rm bh}$--$Q_{\rm jet}$ plane may
provide useful clues to investigations on the central engines in
radio-loud AGNs
\citep*[e.g.,][]{2004MNRAS.349.1419C,2008MNRAS.391..967L}.



In Fig. \ref{histlbol_all}, we show that the distributions of
Eddington ratios for BL Lacs and quasars in our sample exhibit a
bimodal nature. The BL Lac objects are roughly seperated from the
quasars at $L_{\rm bol}/L_{\rm Edd}\sim 0.01$, with most BL Lac
objects having $L_{\rm bol}/L_{\rm Edd}\la 0.01$ and almost all the
quasars having $L_{\rm bol}/L_{\rm Edd}\ga 0.01$. We suggest that
this bimodal behavior of the distribution may imply different
accretion modes in BL Lac objects and quasars, and furthermore the
transition between the accretion states happens at $L_{\rm
bol}/L_{\rm Edd}\sim 0.01$ according to Fig. \ref{histlbol_all}.
Since this is roughly the critical luminosity above which ADAFs are
not possible \citep*[e.g.,][]{1995ApJ...452..710N}, this suggests
that ADAFs  are present in BL Lac objects and standard thin disks
are in quasars. We note that a similar explanation is invoked to
explain the FR I/II division, in which ADAFs would be present in FR
I galaxies while standard thin disks are in FR II galaxies
\citep*[e.g.,][]{2001A&A...379L...1G,2008ApJ...687..156W}.
Interestingly enough, \citet{2004MNRAS.351..733M} found a similar
bimodal distribution of Eddington ratios  for a sample of FR I and
FR II radio galaxies.

As discussed in \S 3, the bolometric luminosities of BL Lac objects
may be over-estimated, if ADAFs are present in these sources. This
would strengthen the bimodality in the distribution of Eddington
ratios of BL Lac objects and quasars.

The similarity between the division of BL Lac objects/quasars and FR
I/II found in this {\it Letter} strongly supports the unification
schemes for FR I/BL Lac object and FR II/radio quasar.

\acknowledgments We thank the anonymous referee for the helpful
comments/suggestions. This work is supported by the NSFC (grants
10778621, 10703003, 10773020, 10821302 and 10833002), the CAS (grant
KJCX2-YW-T03), and the National Basic Research Program of China
(grant 2009CB824800).

{}




\end{document}